\documentclass[amsmath, amssymb, english, aps, prb, floatfix, showpacs,twocolumn]{revtex4}
\usepackage[english]{babel}
\usepackage{graphicx}
\def \mb{\begin{displaymath}} 
\def \me{\end{displaymath}} 
\def \eb{\begin{equation}} 
\def \ee{\end{equation}}   
\def\expect#1{\mathinner{\langle{#1}\rangle}}

{\catcode`\|=\active 
  \gdef\expect#1{\left<\mathcode`\|"8000\let|\bravert {#1}\right>}}
\def\bravert{\egroup\,\vrule\,\bgroup}

\begin{document}

\title{Kondo effect and channel mixing in oscillating molecules}
\author{J. Mravlje$^{1}$ and A. Ram\v{s}ak$^{2,1}$}
\affiliation{$^{1}$Jo\v{z}ef Stefan Institute, Ljubljana, Slovenia}
\affiliation{$^{2}$Faculty of Mathematics and Physics, University of Ljubljana,
Slovenia}

\begin{abstract}
We investigate the electronic transport through a molecule in the
Kondo regime. The tunneling between the electrode and the molecule is
asymmetrically modulated by the oscillations of the molecule, {\it
  i.e.}, if the molecule gets closer to one of the electrodes the
tunneling to that electrode will increase while for the other
electrode it will decrease.  We describe the system by a two-channel
Anderson model with phonon-assisted hybridization. The model is solved
with the Wilson numerical renormalization group method. We present
results for several functional forms of tunneling modulation.  For a
linearized modulation the Kondo screening of the molecular spin is
caused by the even or odd conduction channel. At the critical value of
the electron-phonon coupling an unstable two-channel Kondo fixed point
is found. For a realistic modulation the spin at the molecular orbital
is Kondo screened by the even conduction channel even in the regime of
strong coupling. A universal consequence of the electron-phonon
coupling is the softening of the phonon mode and the related
instability to perturbations that break the left-right symmetry. When
the frequency of oscillations decreases below the magnitude of such
perturbation, the molecule is abruptly attracted to one of the
electrodes. In this regime, the Kondo temperature is enhanced and,
simultaneously, the conductance through the molecule is suppressed.

\end{abstract}


\maketitle

\section{INTRODUCTION}

The Kondo effect, a generic name for processes related to an increased
scattering rate off impurities with internal degrees of freedom,
reveals itself in mesoscopic systems as increased conductance at
biases and temperatures low compared to the Kondo temperature. It has
been observed in measurements of transport through quantum
dots\cite{goldhaber98}, atoms, and
molecules\cite{madhavan98,liang02,park02, yu04_2,
  pasupathy05,zhao05,yu05}. Specific to molecules is the coupling of
electrons to molecular oscillations. The molecular internal
vibrational modes and oscillations of molecules with respect to the
electrodes have been proposed to account for the side-peaks in the
non-linear conductance\cite{park02,yu04_2,pasupathy05}.  In addition,
the electron-phonon coupling can explain the anomalous dependence of
the Kondo temperature on changing the gate voltage at zero bias
\cite{yu05,balseiro06, regueiro07,cornaglia07}. Since the
electrode-molecule junctions are candidates for devices such as
molecular diodes, switches, and rectifiers the research in this field
is increasing despite its complexity and the difficult experimental
characterization \cite{nitzan03,tao06,galperin07}. Recently, also the
notion of quantum phase transition was introduced in the analysis of
such systems\cite{roch08}. We believe that for the interpretation of
the experimental results a better understanding of the behavior of
simple theoretical models is necessary.

\begin{figure}
\begin{center}
\includegraphics[width=40mm, keepaspectratio]{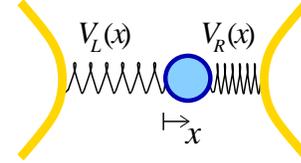}
\end{center}
\caption{\label{Fig1} (Color online) Schematic plot of the model
  device.}
\end{figure}

Here we study the influence of the electron-phonon coupling in the
Kondo regime where a single molecular orbital is occupied on average
by one electron. We concentrate on the case where the oscillations of
the molecule with respect to the electrodes affect the tunneling as
depicted schematically in Fig.~\ref{Fig1}. That is, the tunneling
toward the left and right electrodes is given by overlap
integrals $V_{L,R}(x)$ that are modulated by the displacement $x$ of the
molecule from the mid-point between the electrodes {\it
  asymmetrically}: $V_R$ increases and $V_L$ decreases for $x$
positive, and opposite for $x$ negative.

Assuming the electrodes are identical, it is convenient to introduce
symmetric and antisymmetric combinations of states in the electrodes. With
respect to inversion, they form even and odd conduction channels.  The
odd channel is coupled to the molecule only due to the asymmetric
modulation of tunneling. For example, in the linear
approximation $V_{L,R}(x)=V(1 \mp g x )$ -- a prefactor $V$ sets the
intensity of the tunneling and the electron-phonon coupling constant
$g$ its variation due to the displacement\cite{barisic70} -- the even
channel is coupled to the molecule directly and the odd channel is
coupled to the molecule via a term proportional to $g x$.

As a consequence of coupling the molecular orbital to two channels the
low-energy behavior is that of the two-channel Kondo (2CK)
model\cite{nozieres80,balseiro06}.  The screening of the spin occurs
in the channel with the larger coupling constant. If the couplings
match, an overscreened, {\it i.e.}, a genuine 2CK problem with a
non-Fermi liquid behaviour results. For a linearized model such a
fixed point has indeed been found with simulations based on numerical
renormalization group\cite{mravlje07}. 

 Of interest is also the renormalization of phonon frequencies. Quite
 generally, the characteristic frequency of the oscillations decreases
 with increasing electron-phonon coupling. In the Anderson-Holstein model
 the softening of the phonon mode is related to the increased charge
 susceptibility \cite{hewson02, mravlje05}, which occurs due to 
 dynamical breaking of the particle-hole symmetry for 
 negative effective repulsion $U$. In the present case the softening occurs as
 well and it is related to the dynamical breaking of inversion
 symmetry\cite{balseiro06}. Due to the softening, the instability
 towards perturbations breaking the symmetry emerges. On the mean
 field level\cite{mravlje06}, the instability is seen as an
 asymmetric ground state with large average $x$ in systems
 {\it with} inversion symmetry.

In this work we extend the existing analysis in two ways. (i)
Motivated by the lack of the inversion symmetry of typical
experimental devices we include the inversion symmetry breaking
perturbation.  (ii) We check which features persist if the tunneling
is taken to depend exponentially on the displacement:
$V_{L,R} (x)\propto \exp(\mp gx)$. In particular, it is shown that the
softening of the phonon mode and the corresponding instability occur
universally but the 2CK fixed point appears as an artifact of the
linearization only. It is shown also that the softening is due to the
kinetic energy gained by the dynamical breaking of inversion symmetry
and that it occurs also for vanishing repulsion, $U\to 0$. 

Both directions of research have been pursued in the context of
nano-electromechanical systems\cite{novotny04, twamley06,johansson08},
where only the lowest orders in tunneling are considered and the Kondo
correlations are thus lost. A similar approach has been followed in
analyzing the influence of pair tunneling for negative effective $U$
in the Anderson-Holstein model\cite{koch06,hwang07}.  On the other
hand, the influence of the exponential dependence of tunneling rates
on $x$ in the Kondo regime has been analyzed in
Ref.~\onlinecite{kiselev06}.  However, the displacement $x$ is not
treated as a dynamical variable there but only as an external control
parameter.

The paper is  organized as follows. In the next section we
describe in more detail the models under consideration. We have
performed the numerical 
calculations using Wilson's numerical renormalization group (NRG) 
 and projection operator method of Sch\"onhammer and Gunnarsson (SG)
 which we briefly describe in Section III.  In Section IV we present 
 analytical and in Section V numerical results. We conclude by 
 critically commenting the obtained results and their
 applicability. A comparison between the NRG and SG results is given
 in  Appendix A followed by Appendices B and C containing the derivations
 of  the Schrieffer-Wolff  transformation and the conductance
 formulas. 

\section{Models}
We model the system with the Hamiltonian \eb \label{eq:hami} H=
H_\mathrm{mol} +H_L +H_R + H_{\mathrm{vib}} + H', \ee where
$H_{\mathrm{mol}}$ describes an isolated molecule, $H_L$ and $H_R$
the left and the right electrode, respectively, $H_{\mathrm{vib}}$ a
vibrational mode and $H'$ the phonon-assisted coupling of the
molecular orbital to the electrodes.  The molecule consists of a single
orbital with energy $\epsilon$, which is in experiment modulated by
the gate voltage. The repulsion between two electrons
simultaneously occupying the orbital is $U$,
\begin{equation} 
H_{\mathrm{mol}}= \epsilon (n_\uparrow + n_\downarrow) + U n_\uparrow n_\downarrow, 
\end{equation}
where the number operators $n_\sigma=d_\sigma^\dagger d_\sigma $ count
the number of electrons in 
the orbital with spin $\sigma=\uparrow,\downarrow$. The symbols
$c^{(\dagger)},d^{(\dagger)}$  denote electron annihilation
(creation) operators in the electrodes and molecular orbital,
respectively. We are here interested in the particle-hole symmetric
point $\epsilon=-U/2$ only. The oscillator
part is 
\eb H_{\mathrm{vib}} = \Omega a^\dagger a, \ee
describing the oscillations with frequency $\Omega$ and
$a^{\dagger}$ is the boson creation operator. 
 The left and right electrodes
are described by bands of noninteracting electrons $H_{\alpha} = \sum_{k \sigma}
\epsilon_k n_{k \alpha\sigma} $ for $\alpha=L,R$, respectively, 
where $n_{k \alpha \sigma} =c^\dagger_{k \alpha \sigma} 
c_{k \alpha \sigma}$ counts the electrons with spin
$\sigma$ 
and wave vector $k$;
 $\epsilon_k$ is the dispersion of the band in the electrode
$\alpha$.  The chemical potential is set to the middle of  the  band
($\mu=0$) corresponding to the half-filled regime where the molecule is
on average singly occupied.  In NRG calculations a flat band with
constant density of states $\rho=1/(2D)$ 
and in SG calculations a  tight-binding band with $\rho(\omega)
=1/\left(\pi \sqrt{D^2-\omega^2}\right)$ are used, where $D$ is the half-width
  of the band.

 The tunneling between the molecular orbital and the electrodes, which is
 described by \eb H'= V_L(x) \widehat{v}_L + V_R(x) \widehat{v}_R, \ee
 occurs via the hybridization operators (assuming here the tunneling
 is $k$-independent) \eb \label{eq:hyb}
 \widehat{v}_{\alpha}=\sum_{k\sigma} c_{k\alpha\sigma}^{\dagger}
 d_\sigma + h.c.  \ee multiplied by the overlap integrals
 $V_{\alpha}(x)\equiv V_\alpha(a+a^{\dagger})$, where the displacement is
 explicitly quantized.

It is practical to define even and odd combinations of states in the
electrodes, respectively

\eb c^\dagger_{ke(o)\sigma} =\frac{1}{\sqrt{2}} \left(c^\dagger_{kL\sigma} \pm c^\dagger_{kR
\sigma} \right). \ee 
In this basis $H'$ reads 
\eb H'=   V_e(x)\widehat{v}_e + V_o(x) \widehat{v}_o\ee
where  
\eb V_{e,o} (x)=\frac{V_L(x)\pm V_R(x)}{\sqrt{2}}, \ee
modulate the tunneling to even and odd channels. Hybridization operators $\widehat{v}_{e,o}$
  correspond to Eq.~(\ref{eq:hyb}) for $\alpha=e,o$,
  respectively. Note that \eb\label{eq:ineq}|V_e(x)| >|V_o(x)|\ee if
$V_{L,R}(x)$ are both positive or both negative for all $x$.

In this paper we perform the calculations  using several 
functional forms of $V_\alpha(x)$ depicted in Fig.~\ref{Fig_func}.

\subsection{Overlap integrals}
In a realistic experimental situation the tunneling between the
molecule and the tip of an electrode will be saturated at small
distances and it will progressively decrease with increasing distance of
the molecule from the electrode.  The precise functional dependence
of overlap integrals will in general depend on details of the molecule
and the tips of the electrodes, but the overall behavior should be as
shown in Fig.~\ref{Fig_func}(a) with dotted line.

\subsubsection {Linear modulation }
The simplest form of overlap integrals is obtained by the expansion to
lowest order in displacement resulting in linear modulation (LM) \eb
V_{L,R} (x)=V\left[1\mp(gx +\zeta)\right]. \ee The tunneling matrix
element, constant $V$ for $g=0$, is linearly modulated by displacement
for $g > 0$. We assume the system is almost inversion symmetric. A
small $\zeta \geq 0$ is the magnitude of the symmetry breaking
perturbation. In the symmetrized basis the overlap integrals take on
the following form \eb\label{eq:linmod} V_e=\sqrt{2} V, \; \;
V_o=\sqrt{2}V (gx + \zeta).\ee Note that Eq.~(\ref{eq:linmod}) does
not satisfy the requirement Eq.~(\ref{eq:ineq}) for $gx> 1-\zeta$,
because  the overlap to the left electrode becomes negative and
its absolute value starts to increase with increasing $x$ (dashed
region in Fig.~\ref{Fig_func}).

\subsubsection{ Exponential modulation}

A better approximation to the overlap integrals could be exponential
decay at large distances and arguably more realistic
model is given with the exponential modulation (EM) of tunneling by
\eb V_{L,R}(x)=V \left[\exp(\mp gx) \mp \zeta\right], \ee or
equivalently

\eb V_e=\sqrt{2} V\cosh(gx),\: \: V_o =\sqrt{2} V
\left[\sinh(gx)+\zeta\right],\ee which are positive and for $\zeta=0$
manifestly satisfy the relation Eq.~(\ref{eq:ineq}).   By expanding the
couplings to lowest order in $x$, the LM is recovered.

\begin{figure}
\begin{center}
\includegraphics[width=65mm, keepaspectratio]{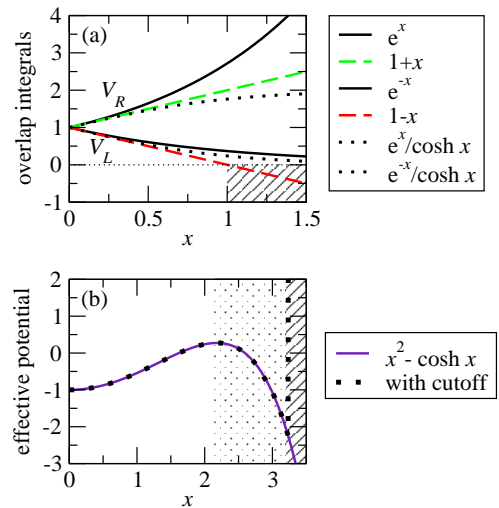}
\end{center}
\caption{\label{Fig_func} (Color online) (a) Various forms of the
  tunneling-modulation.  The unphysical regime of LM where the
  tunneling starts to increase with increasing distance to the
  electrode is indicated by dashing. (b) The breakdown of EM. In the
  right-hand part the effective potential drops without bounds. This
  is regularized by the phonon cutoff, which corresponds to the
  hard-wall boundary. Two different cutoff regimes are indicated by
  dotted and dashed area.}
\end{figure}

The EM eliminates the negative overlap but introduces another problem
due to divergence of $\exp(gx)$ at large $gx$. Namely, the model with
EM is unstable towards large displacements as can be understood by the
following simple argument. For large $x$, the largest energies in the
problem are $V \exp(gx)$ and $\Omega x^2$. This limit corresponds to
two sites coupled by a tunneling term with tunneling proportional to
$\exp(gx)$. The total energy of one electron on these two sites is
$E\sim \Omega x^2/4-V \exp(gx) $. Therefore the oscillator in the EM
moves in an effective potential of the form depicted in
Fig.~\ref{Fig_func}(b) (full), unbounded from below for large $|x|$.

\subsection{Limitations of models}

In real systems, the overlap integrals will be neither negative nor
divergent.  The large $x$ behaviour of EM can be corrected by adding
higher terms in displacement to the oscillator potential, which
corresponds to hardening of the 'spring' for large $x$. In our
numerical calculations such a hardening is incorporated in the form of
a phonon cutoff which acts as a hard-wall boundary, thereby
eliminating the states corresponding to displacements larger than
$\sim 2\sqrt{L}$, where $L$ is the maximal number of phonons
allowed. In Fig.~\ref{Fig_func}(b), the dotted and the dashed regions
indicate two different cutoff regimes. For the larger cutoff also the
resulting effective potential is sketched (dotted).

By incorporating the cutoff into EM the results qualitatively depend
on additional parameter $L$ because the choice of cutoff determines
the form of the effective potential near low energies. However,
without some kind of a regularization of the model the model with EM
is ill defined; we show later that the average displacement (or its
fluctuations for $\zeta=0$) diverge for all $g>0$ at $L\to \infty$.

We note that there may be several cases, where the modulation is not a
simple function.  To describe the experimental situation, it might
even be needed to include the anharmonicity of the potential as well.
However, a convenient starting point is to first clarify specific
regimes of simplified models and the consequences of the
approximations.  In this paper, we first analyze the model with LM
comprehensively. Later we discuss EM for a specific phonon cutoff to
highlight which of the results obtained using LM are artifacts of the
linearization. Finally, the exponential divergence of overlap
integrals is regularized, Fig.~\ref{Fig_func}(a) (dotted), and it is
shown which results persist also for this model.

\section{Numerical methods} 
Most of the numerical results presented here have been obtained using the 
 Wilson
numerical renormalization group\cite{wilson75,bulla08}(NRG)
method. The NRG procedure
is based on adding sites to the system iteratively with hopping matrix
element to
the $n$th added site decreasing as $\Lambda^{-n/2}$. At each step the
resulting   Hamiltonian is diagonalized  and lowest $K$ eigenstates are
kept. The exponentially decreasing hopping is essential to introduce all
the energy scales while still keeping the numerical effort
reasonable. Such a procedure is especially suitable for the Kondo problem
where a range of energy scales contributes equally to the screening of
the impurity spin. The algorithm is stopped after $N_\mathrm{max}$
iterations. In the presented results we have typically used $\Lambda=3-4$,
$K=2000$ (not counting the degeneracies due to spin, isospin, and
parity symmetries\cite{zitko06a} which have been explicitly taken into
account) and $N_{\mathrm{max}}=40$.

In order to gain additional insight and  to make a relation with
our previous work we compare the NRG results to the results obtained by
the  Sch\"onhammer-Gunnarsson (SG) \cite{schonhammer76,gunnarsson85}
variational method. The details of our implementation of the 
variational method are given in our previous work \cite{rejec03b,
  rejec03a,mravlje05}.  For reader's 
convenience we here just remark that it consists of finding the
parameters of an auxiliary noninteracting Hamiltonian $\tilde{H}$ [of the same 
form as $H$ in Eq.~(\ref{eq:hami}), but for $g=0, U=0$ and described
by renormalized parameters
$\tilde{V}_L,\tilde{V}_R,\tilde{\epsilon}$], which minimize the 
variational ground state energy $E=\langle\Psi | H|\Psi\rangle$, 
where the variational function $\Psi$ is expressed in the basis of
projection operators $P_{i}$ acting on the 
Hartree-Fock ground state  $|\Psi_0 \rangle$ (which includes the
phonon vacuum) of the auxiliary
Hamiltonian $\tilde{H}$,
\begin{equation} |\Psi\rangle = \sum_{ni} \psi_{ni} (a^{\dagger})^n P_i |\Psi_0
  \rangle.\end{equation}

We have adapted the SG method also to extract the effective oscillator
potential. By restricting the parameters of the auxiliary Hamiltonian
(for example, by fixing $\tilde{V}_L/\tilde{V}_R=r$), the minimization procedure gives
states $|\Psi_r\rangle$, for which the $\langle \Psi_r | \widehat{x}
|\Psi_r \rangle =x_r$ is in general finite, and energies $E_r =\langle
\Psi_r | H |\Psi_r \rangle$. The effective potential is estimated by
pairs ($x_r$,$E_r$).

\section{Analytical results}
We studied the model Eq.~(\ref{eq:hami}) numerically and the results
are presented in Section V.  Nevertheless, from analytical results in
special limits we anticipate different regimes of behavior and the
values of parameters where these regimes emerge.
\subsection{Linear modulation}

For $U$ and $\Omega$ large the low energy behaviour is obtained by
projecting the Hamiltonian onto space consisting of states with singly
occupied molecular orbital and without excited phonons.  The result of
this Schrieffer-Wolff (SW) transformation (described in Appendix B) is
the 2CK Hamiltonian \eb H_\mathrm{2CK}=J_e \mathbf{S} \cdot
\mathbf{s}_e + J_o \mathbf{S} \cdot \mathbf{s}_o, \ee describing the
anti-ferromagnetic coupling, between the spin on the molecular orbital
$\mathbf{S}$ and the spin densities $\mathbf{s}_\alpha$ in orbitals
next to impurity in the even and odd channel, $\alpha=e,o$,
respectively. The coupling constants are
\begin{equation}  J_e = 2 V_e^2 \left(
    \frac{1}{-{\epsilon}}  +
  \frac{1}{{\epsilon} +U}\right)\end{equation}
and \begin{equation} 
J_o=2 V_o^2 \left(\frac{1}{-\epsilon + \Omega} + \frac{1}{\epsilon+ U
  + \Omega} \right). \end{equation}
The ratio between the coupling constants
\eb \frac{J_o}{J_e}=\frac{g^2}{1+2
\Omega/U} \label{eq:ration} \ee 
determines in  which of the two channels the Kondo screening takes
place. Let $g_\mathrm{c}$ denote the delimiting value separating regimes
with different symmetries of the screening channel. At $g_c$ both
channels participate equally to the screening and the over-screened non-Fermi
liquid behavior results. In terms of the
original model, this corresponds to the point at which
inequality Eq.~(\ref{eq:ineq}) is violated. According to
Eq.~(\ref{eq:ration}), for large $U, \Omega$, 
\begin{equation} g_\mathrm{c} \sim 
\sqrt{1+2\Omega/U}.\end{equation}

\begin{figure}
\begin{center}
\includegraphics[width=73mm, keepaspectratio]{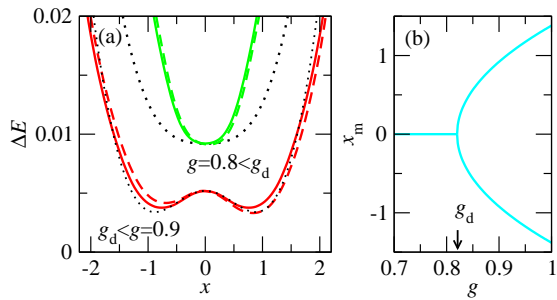}
\end{center}
\caption{\label{Fig_pot_simple} (Color online) (a) Effective oscillator
  potential for LM.  Semi-classical estimate (dotted); SG estimate for
  $\zeta=0$ (full) and $\zeta=0.01$ (dashed). Parameters
  $\Omega=0.1$, $\Gamma=0.02$ and $U=0.3$ (for SG only) are in units
  of $D$ (half-width of the  band). (b) The minima of the potential in
semi-classical estimate.} 
\end{figure}

We now turn to the renormalization of the vibrational mode and
demonstrate that through the electron-phonon coupling the confining
potential is diminished and can even be driven to the form of a double
well. We first discuss $U=0$ model and use the semi-classical
approximation in which phonon operators are substituted, $a \sim
a^\dagger \to x/2$, by a real valued constant $x$.  The resulting
Hamiltonian with the hybridization
$\tilde{\Gamma}(x)=\Gamma(1+g^2x^2)$ is noninteracting and can be thus
solved exactly. Here the magnitude of the bare hybridization with
$g=0$ is defined by $\Gamma=\pi V^2/D$. In this model only the
hybridization energy gain, which in the wide-band limit ($\Gamma/D$
small) reads\cite{fabrizio07} $\Delta E_{\mathrm{hyb}}=-2/\pi
\tilde{\Gamma} \log D/\tilde{\Gamma}$, and the elastic energy cost
$\Delta E_{\mathrm{el}}=\Omega x^2/4$ are a function of $x$. Hence,
the effective oscillator potential in this estimate is $\Delta
E_\mathrm{SC}=\Delta E_{\mathrm{el}}+ \Delta E_{\mathrm{hyb}}$ and can
be written in a closed form
\begin{equation} \Delta
E_\mathrm{SC}(x)=\Omega x^2/4 -(2/\pi) \tilde{\Gamma}(x)  \log
  \{D/[\tilde \Gamma (x)]\}.
  \label{eq:enerSC}
\end{equation}
 The prefactor of the $x^2$-term in the small-$x$ expansion is equal
 to $\Omega/4 - \left\{(2/\pi) g^2 \Gamma \left[\log(D/\Gamma)
   -1\right]\right\}$ and is decreasing with increasing $g$ indicating
 the softening of the confining potential. At sufficiently large $g$, 
two wells emerge. In
Fig.~\ref{Fig_pot_simple} (a) we plot (dotted) the resulting potential
for $g$ below and above the delimiting value ($g_\mathrm{d}$). The
$g_\mathrm{d}$ and the 
positions $x_\mathrm{m}$ [shown in Fig.~\ref{Fig_pot_simple}(b)] of
the potential minima can be  extracted analytically from  
 Eq.~(\ref{eq:enerSC}),
 \begin{eqnarray} 
g_\mathrm{d} &=& \sqrt{\frac{\pi\Omega}{8 \Gamma
    \left[\log(D/\Gamma)-1\right]}} ,\label{eq:semiclass}\\
    x_\mathrm{m} &=& \sqrt{\frac{\pi\Omega(g-g_\mathrm{d})}{4 \Gamma
   g_\mathrm{d}^5}}.
\label{eq:semiclassx}
\end{eqnarray}

For a finite $U$ the model cannot be solved exactly, but by estimating
the potential with the SG method, we find that the evolution of the
potential as just described persists.  The SG results for $U>0$  are plotted
in Fig.~\ref{Fig_pot_simple} for $\zeta=0$ (full
lines) and also for the case with broken left-right symmetry
$\zeta=0.01$ (dashed). While for $g<g_\mathrm{d}$ the potential is only
slightly perturbed, finite $\zeta$ for $g>g_\mathrm{d}$ breaks the
degeneracy between the two minima. In this regime, the molecule is
attracted to one of the electrodes.

Having defined the characteristic values $g_\mathrm{c}$ and
$g_\mathrm{d}$ it is interesting to ask whether there is any relation between
the two. As discussed above, the 2CK  point occurs at $g_\mathrm{c}$
such that in the semi-classical description $1-g_\mathrm{c}|x|
\sim0$, meaning that the double well-potential has to be
preformed for its minima to occur at $ |x_\mathrm{m}| \gtrsim
1/g_\mathrm{c}$. Therefore we expect
$g_\mathrm{c}>g_\mathrm{d}$. However, as  $x_\mathrm{m}$ evolves
rapidly as a function of $g$ the values
of $g_\mathrm{c}$ and $g_\mathrm{d}$ are close.

To support this statements quantitatively in Table~\ref{table_SW} we
show the semi-classical (SC) and SG estimates of $g_\mathrm{d}$ compared
to the SW and NRG estimates of $g_\mathrm{c}$ for several values of
parameters.  As expected, we  find that the
Schrieffer-Wolff estimate of $g_\mathrm{c}$ becomes more accurate (agrees
better with the NRG) for $U,
\Omega$ large when the charge fluctuations and phonon excitations are
suppressed.  Conversely, the semi-classical estimate is more accurate
for 
large number of excited phonons  (small $\Omega$) and small $U$ but
breaks down for large $U$. For example, for $\Omega=1$ and $U=3$ the
value obtained by  
the semi-classical method  overestimates $g_\mathrm{d}$ to a
value which is larger than $g_\mathrm{c}$ obtained using NRG.

\begin{table}%
\caption{\label{table_SW}  SC and SG: $g_\mathrm{d}$ obtained 
  from semi-classical (SC) estimate and  from
  SG simulations. SW and NRG: $g_\mathrm{c}$   calculated by
  SW estimates and from NRG simulations. 
  SG method is inapplicable for very large $U$.}
\begin{ruledtabular}
\begin{tabular}{r r r r r r}
$\Omega$ & $U$ & SC & SG & SW & NRG\\
\hline
0.1 & 0.3 & 0.82 & 0.85  & 1.29 & 0.84 \\
0.1 & 0.6 & 0.82 & /  & 1.15 & 0.85   \\
1   & 0.03 & 2.60 & 2.59  &  8.22 & 3.11\\
1 & 0.3 & 2.60 & 2.32 & 2.76 & 2.32   \\
1 & 3 & 2.60 & / & 1.29 & 1.28   \\
\end{tabular}
\end{ruledtabular}
\end{table}

\subsection{Exponential modulation}
Let us perform the Schrieffer-Wolff transformation  also on the
model with exponential modulation of tunneling. We obtain 
\eb  J_{e,o}=2 V^2 \sum_{m=0}^\infty \left (\frac{\delta_m^{e,o}}{-\epsilon +m \Omega}
+ \frac{\delta_m^{e,o}}{\epsilon + U + m \Omega} \right) \ee
where $\delta_m^e=|\langle 0 | \cosh (gx) |m\rangle|^2$ and
$\delta_m^o=|\langle 0 | \sinh (gx) |m\rangle|^2$. Note that as $J_e>J_o$
for all values of parameters no 2CK fixed point occurs in such
a model. Note also that both $J_\alpha$ depend on $g$ exponentially. The Kondo temperature, which itself is exponential in $J$, $T_K
\propto \exp(-1/\rho J)$, is thus very sensitive to the value of $g$.

  The effective potential which is unbounded from below for $L\to
  \infty$ is regularized by the phonon cutoff $L$. At fixed $L$ we
  define $g_\mathrm{e} (L)$ as the value of $g$ at which the molecule
  is attracted by the hard wall boundary as signaled by an abrupt
  increase of displacement (shown later).  Given $L$, the behavior of
  the model for $g> g_\mathrm{e}(L) $ becomes dominated by the
  hard-wall boundary only.  For instance, the full curve in
  Fig.~\ref{Fig_func}(b) corresponds to a particular $g$ which is
  smaller than $g_\mathrm{e}$ for the smaller phonon cutoff (dotted
  region) and larger than $g_\mathrm{e}$ for the larger phonon cutoff
  (dashed region).  For $L\to \infty$, $g_\mathrm{e}\to 0$.  %

\section{Numerical results}
In this section we confirm  the anticipations stated above with
numerical examples. We first show the results for LM and then for EM. We
treat separately the inversion symmetric $\zeta=0$ and asymmetric
$\zeta>0$ cases. We
use the half-width of the band $D$  as the energy unit. Unless where
 explicitly stated, we take 
$U=0.3$, $\Gamma=\pi V^2/D=0.02$, and $\Omega=0.1$. All the results
correspond to the particle-hole symmetric point,
$\epsilon=-U/2$, and to the zero temperature limit.

\begin{figure}[h]
\begin{center}
\includegraphics[width=60mm, keepaspectratio]{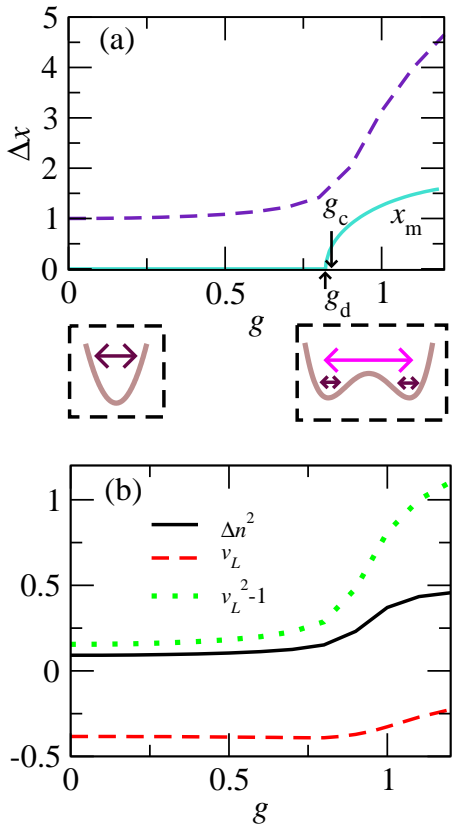}
\end{center}
\caption{\label{Fig2} (Color online) (a) Fluctuations of
  displacement. (b) Fluctuations of charge, hopping to one of the
  electrodes and its fluctuations. 
 The left (right)
  pictograms schematically present the effective oscillator potential
  before (after) the emergence of the soft mode ($U=0.3, \Gamma=0.02,
    \Omega=0.1$).}
\end{figure}

\subsection{Linearized model}
\subsubsection{Inversion symmetry: $\zeta=0$}

We begin by looking at the static quantities for $\zeta=0$. The
average displacement $\langle x \rangle $ which is odd under inversion
vanishes.  The fluctuations of displacement $\Delta
x=\langle (x-\langle x \rangle )^2\rangle^{1/2}$, shown in
Fig.~\ref{Fig2}(a), increase monotonically with $g$. The slope of
$\Delta x$ increases considerably at $g\sim g_\mathrm{d}$
$(\sim g_\mathrm{c})$, where the double well  effective potential
is formed, as indicated in pictograms.  The change in slope is driven
by the increased hybridization in the odd-channel. The position of
the minima of effective potential $x_\mathrm{m}$ (full line)
Eq.~(\ref{eq:semiclassx}) also becomes non-vanishing there. 

Due to increased hybridization the fluctuations of charge $\Delta
n^2=\langle (n - 1)^2 \rangle$, shown in Fig.~\ref{Fig2}(b), are
increased in the $g>g_\mathrm{c}$ regime. However, the absolute value
of the average of the hybridization operator $v_L=\langle
\widehat{v}_L \rangle=\langle \widehat{v}_R \rangle$ is diminished,
contrary to the naive expectation. This is due to increasingly
fluctuating sign of the overlap integral in this regime.  On the other
hand, the average of the hybridization operator squared $v_L^2=\langle
\widehat{v}_L^2 \rangle$ is increased here, as expected.

\begin{figure}
\begin{center}
\includegraphics[width=78mm, keepaspectratio]{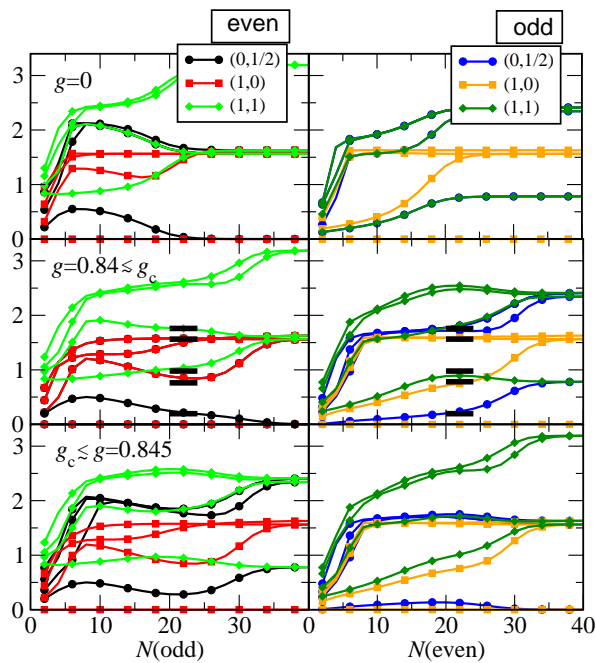}
\end{center}
\caption{\label{Fig3} (Color online) NRG flow diagram.  In the left (right)
   panels eigen-energies of states of even (odd) parity are shown for
   odd (even) number of NRG iterations, 
   respectively. The states are labeled by ($Q$,$S$): total charge $Q$
   and total spin $S$ quantum numbers. 
  The thick
   horizontal bars are the conformal field    theory predictions for
   2CK fixed point.  Parameters are as
  in Fig.~\ref{Fig2}.
  }
\end{figure}

In Fig.~\ref{Fig3} we plot the NRG flow diagram: the energies of the
lowest few eigenstates in units of characteristic energy of a
particular iteration $\omega_N \propto \Lambda^{-(N-1)/2}$ as the
function of the NRG iteration number $N$. The fingerprint of the Fermi
liquid ground state are the equidistant low-lying quasiparticle
excitations\cite{affleck92}, which are seen for large $N$,
irrespective of $g$. By comparing the top two panels with the bottom
panel it is seen, that the roles of even and odd parity states are
interchanged in the Fermi-liquid regime (right-hand side of each
panel) corresponding to the change of the screening channel as $g$ is
increased above $g_\mathrm{c}$.

 For $g\sim g_{\mathrm{c}}$ (bottom panels) the unstable non-Fermi
 liquid fixed point, which determines the NRG flow at intermediate $N$
 ($\sim 20$ for the plotted case) is discerned.  The ratios of
 eigen-energies here (horizontal bars) are characteristic of the 2CK
 effect and agree with the predictions of conformal field theory
 \cite{affleck92} $(0,1/8,1/2,5/8,1,9/8,...)$. This regime cannot be
 explained in terms of the Fermi liquid quasiparticles. The
 difference between the couplings to the screening channels is a relevant
 perturbation and for low temperatures (large $N$) drives the flow
 towards the (stable) Fermi liquid fixed point.

Now we turn to the renormalization of the phonon propagator by the
electron-phonon coupling. The dynamical information about oscillator is contained in the
displacement Green's function. The displacement spectral function 
\begin{align} \mathcal{A}(\omega)&=-\frac{1}{\pi} \mathrm{Im} \ll x,x \gg_\omega
  = \nonumber \\ &=-\frac{1}{\pi}\mathrm{Im} \int_{0}^{\infty} (-i)
  \langle[x(t),x(0)]\rangle e^{i \omega t} dt \end{align} is an odd
function of $\omega$ due to the hermiticity of $x$ (unlike $\mathrm{Im}\ll a, a^\dagger \gg_\omega$ which is odd only 
  for the inversion symmetric case $\zeta=0$).  Since in NRG 
$\mathcal{A}(\omega)$ is  evaluated for a finite system it
consists of several $\delta$-peaks of different weights. To obtain a
smooth spectral function we have used the Gaussian broadening on the
logarithmic scale\cite{bulla01}, where the Dirac $\delta$ function is
broadened according to %
\eb \delta(\omega - \omega_n) \to \frac{1}{b \omega_n \pi} \exp \left\{ -\left[\frac{\log
    (\omega/\omega_n)}{b}\right]^2 -\frac{b^2}{4}\right\}, \ee and we
used $b=0.3$ in our calculations.

\begin{figure}
\begin{center}
\includegraphics[width=68mm, keepaspectratio]{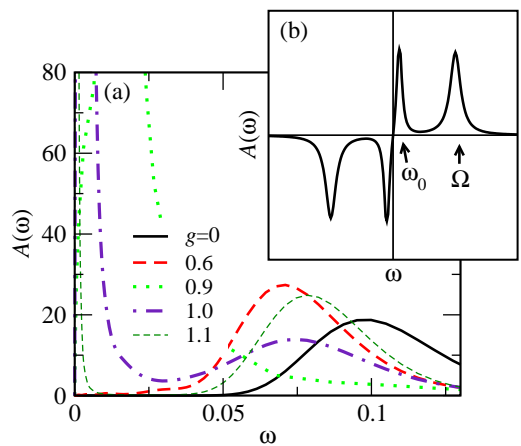}
\end{center}
\caption{\label{Fig4} (Color online) (a) Displacement spectral functions
  for various $g$.  Note the 
  softening of the phonon mode. For large $g$ only small
  amount of spectral weight resides at low frequencies. The peaks at
  large frequencies appear broader than they should be because of the
  broadening procedure described in the text. Parameters are as
  in Fig.~\ref{Fig2}. (b) Spectral function after the emergence of the
  soft mode schematically. The frequencies of the soft mode $\omega_0$
  and the high-frequency oscillation at $\omega \sim \Omega$ are
  indicated. 
}
\end{figure}

In Fig.~\ref{Fig4}(a) we plot $\mathcal{A}(\omega)$ for various $g$. The width of
the high frequency peaks is overestimated (the extreme example is the
$g=0$ peak at $\Omega$ for which the width should vanish) due to the
broadening procedure described above.     We could
use Dyson equation \cite{bulla98, jeon03} to obtain sharper peaks
but on one hand there is no {\it a priori} guarantee that such a
procedure gives more accurate results for large $g$ 
and on the other hand we do not use the width of the peaks as a means
to draw any quantitative conclusion.

The evolution of the phonon operator can be understood in terms of the
evolution of effective confining potential (Fig.~\ref{Fig_pot_simple},
pictograms in Fig.~\ref{Fig2}). For intermediate $g$ [starting at
  $g\sim 0.5$ for the parameters used in Fig.~\ref{Fig4}(a)] as the
confining potential starts to diminish the vibrational mode begins to
soften: the peak of $\mathcal{A}(\omega)$ moves to lower
frequencies. At still larger $g\sim g_\mathrm{d}$ two peaks emerge at
the point where the two wells develop in the effective
potential. Characteristic dependence of $\mathcal{A}(\omega)$ in this
regime is schematically presented in Fig.~\ref{Fig4}(b).  The double
well potential is already well established and the major part of the
spectral weight corresponds to the oscillations within each of the
wells. The minor, low frequency part corresponds to increasingly slow
tunneling (see also Fig.~\ref{Fig7}) between the degenerate minima of
the potential. The frequency $\omega_0$ and the weight of the low
frequency peak decrease with increasing $g$.

Due to the inversion symmetry the conductance can be calculated using
the scattering phase shifts only, as described in Appendix C. The
scattering phase shifts in the even and odd channels are $\pi/2 (0)$
and $0(\pi/2)$, for $g<g_\mathrm{c}$ $(g>g_\mathrm{c})$ and the
conductance evaluated by Eq.~(\ref{eq:cond_phase_shift}) is unity
\cite{mravlje06}. 

\subsubsection{Broken inversion symmetry:  $\zeta>0$} 
It is impossible to experimentally produce perfectly symmetric
devices, therefore it is interesting to check for the influence of the
inversion symmetry breaking term of relative strength $\zeta$. Let us
first remark that the NRG flow diagrams (not shown here) are that of the
Fermi liquid as the occurrence of 2CK fixed point is inhibited by the
breaking of inversion symmetry.

\begin{figure}
\begin{center}
\includegraphics[width=52mm, keepaspectratio]{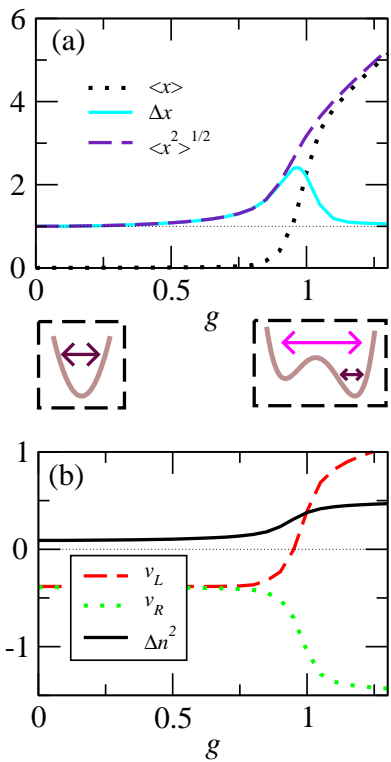}
\end{center}
\caption{\label{Fig5} (Color online) Static quantities for system with
  broken inversion symmetry. (a) Averages of $x$, $x^2$, and $\Delta
  x$. (b) Averages of hopping $v_L$ and $v_R$ and fluctuations of
  charge.  The breaking of inversion symmetry $\zeta=0.01$. Other
  parameters as in Fig.~\ref{Fig2}.  }
\end{figure}

In Fig.~\ref{Fig5} we plot static correlations for $\zeta=0.01$.  New
compared to the inversion symmetric case is the non-vanishing average
displacement, which monotonically increases with $g$. Despite the
simultaneous increase in average of $x^2$, the fluctuations of
displacement $\Delta x=\langle(x-\langle x \rangle)^2\rangle^{1/2}$
eventually reach a maximum. At a still larger $g$ the molecule is
attracted to the right electrode as indicated in the right pictogram. The
fluctuations of charge are remain the same as in the $\zeta=0$ case
but the average hopping to the right electrode $v_R$ is larger than $v_L$,
which first vanishes and then changes sign. This asymmetry happens not
at $g_\mathrm{c} \sim g_\mathrm{d}$ but at another value
$g_\mathrm{asy}$ ($\sim 1$ for these parameters), when the softened
phonon frequency $\omega_0$ and the energy difference between the
hybridizations to left and right $\propto \zeta$ become comparable.

\begin{figure}
\begin{center}
\includegraphics[width=62mm, keepaspectratio]{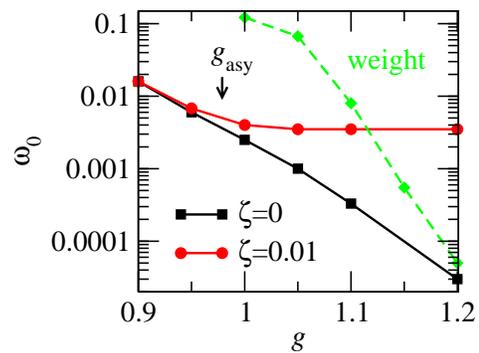}
\end{center}
\caption{\label{Fig7} (Color online) The frequency (full lines) of the
  soft mode peak as a function of $g$.  The weight of the soft mode
  peak for $\zeta=0.01$ and normalized to some arbitrary value
  (dashed). Other parameters are as in Fig.~\ref{Fig2}.  }
\end{figure}

The spectral functions for $\zeta=0.01$ differ from the $\zeta=0$ case
only for $g>g_\mathrm{asy}$. The distinction between the two cases is
mainly that for $\zeta=0.01$ the frequency of the soft mode
oscillation $\omega_0$ saturates. In Fig.~\ref{Fig7}, we plot the
$\omega_0$ for both $\zeta$ as a function of $g$. For $\zeta=0.01$, we
also plot the weight of the $\omega_0$ peak, which diminishes 
exponentially with increasing $g$.

Again, the behavior is easily understood in terms of the effective
potential pictorially shown in Fig.~\ref{Fig5}. For small to
intermediate $g$ the soft mode begins to emerge as the shape of the
potential is transformed to a double-well-like form with the right
well being lower in energy by a value $\Delta \propto \zeta$. When $g$
increases further and $\omega_0$ decreases below $\Delta$, the
tunneling is suppressed. In this regime, the major part of the
displacement spectral weight is due to the tunneling within the lower
of the wells. The average displacement increases, its fluctuations
decrease. Note that even a minor breaking of inversion symmetry
results in a strongly asymmetric state through the mechanism described
here.

\begin{figure}
\begin{center}
\includegraphics[width=60mm, keepaspectratio]{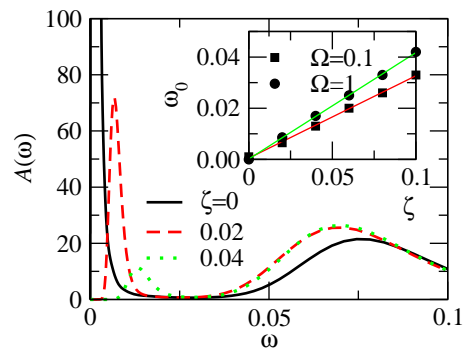}
\end{center}
\caption{\label{Fig8} (Color online) (a) Displacement spectral
  functions for fixed $g=1.05$ and $\zeta=0,0.02,0.04$. Note the
  reduction of weight of low frequency peak on increasing $\zeta$. The
  high energy peak of $\zeta=0$ case is approached at still smaller
  $\zeta$. Inset: The position of low frequency peak as a function of
  $\zeta$ for $\Omega=0.1$, $g=1.05$ and $\Omega=1$, $g=7$. The full
  lines are fits to the data with slopes 0.32 and 0.42 for
  $\Omega=0.1,1$, respectively. Other parameters are as
  in Fig.~\ref{Fig2}.  }
\end{figure}

We plot also $\mathcal{A}(\omega)$ for different $\zeta$ and fixed
$g=1.05>g_\mathrm{asy}$ in Fig.~\ref{Fig8}.  The characteristic
frequency of the soft mode is related to the energy-difference of the
two wells and is proportional to $\zeta$ as shown in the inset of
Fig.~\ref{Fig8}.

\begin{figure}
\begin{center}
\includegraphics[width=60mm, keepaspectratio]{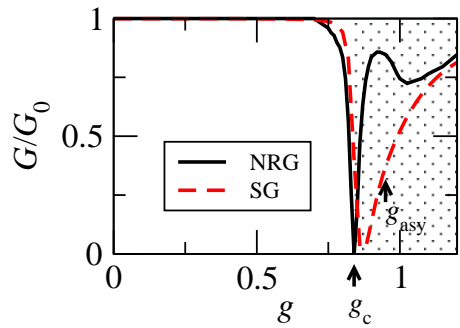}
\end{center}
\caption{\label{Fig_cond} (Color online) Conductance as calculated
  from the current-current correlation function obtained by NRG
  (full) and from the SG method. The shaded region indicates the
  region the results are outside of the scope of the linearized
  model. $\zeta=0.01$, other parameters are as
  in Fig.~\ref{Fig2}.
}
\end{figure}
The conductance for $\zeta>0$ cannot be obtained from the scattering
phase shift alone since the rotation angle $\theta$ (see Appendix C)
is not known in the present case of broken symmetry. Therefore we
calculated the conductance from the current-current correlation
function. We plot the conductance calculated by NRG and SG in
Fig.~\ref{Fig_cond}. The conductance decreases to zero at $g \sim
g_\mathrm{c}$. This minimum corresponds to the virtual decoupling of
the left electrode, $1-gx \sim 0$, there on the average. Due to
linearization, for still larger $g$ the magnitude of the overlap to
the left electrode increases again. Correspondingly, the conductance
is increased. The access to this region (denoted shaded)
is probably unaccomplishable in measurements of transport through
molecules.

The NRG and the SG data agree well for most $g$. However, for $g \sim
g_\mathrm{asy}$ the maximum and minimum are seen in the NRG results, a
feature which the effective Hamiltonian of the SG method fails to
capture.  The discrepancy is especially visible for the
parameters used here for which $g_\mathrm{c} \sim g_\mathrm{asy}$
(compare also with data given in Appendix A). The NRG and SG for
$g>g_\mathrm{asy}$ agree again. Here the fluctuations of displacement
are diminished and the behavior is efficiently described in terms of
the effective Hamiltonian with asymmetric coupling to the electrodes.

\subsection{Exponential model}
\begin{figure}
\begin{center}
\includegraphics[width=65mm, keepaspectratio]{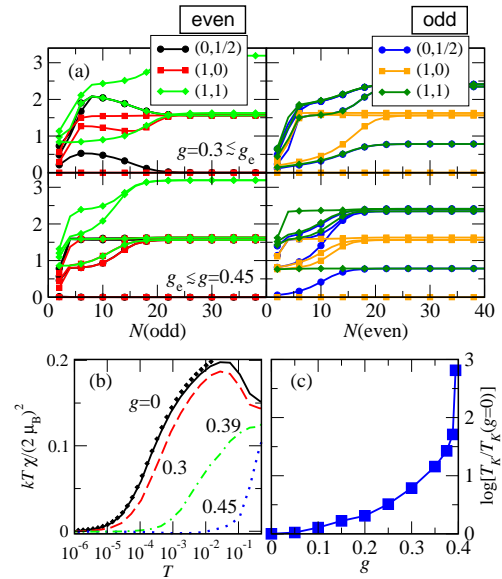}
\end{center}
\caption{\label{exp4} (Color online) Model with exponential modulation
  of overlap integrals (EM). $L=10$, other parameters as in
  Fig.~\ref{Fig2}. (a) NRG eigenvalues. (b) Impurity contribution to
  magnetic susceptibility. For $g>g_\mathrm{e}$ the local moment
  regime is absent.  Diamonds indicate the susceptibility calculated
  using the Bethe ansatz. (c) The Kondo temperature.  }
\end{figure}

We now turn to the model with exponential modulation. Here even for
the inversion symmetric model ($\zeta=0$) the 2CK Kondo fixed point is
inaccessible because the coupling to the even channel is invariably
stronger than the coupling to the odd channel. Correspondingly, the
finite size spectra shown in Fig.~\ref{exp4}(a) are that of the Fermi
liquid ground state with the spin screened by the even conduction
channel.  Even at large $g$, the even and the odd channels do not
inter-change their roles in the screening, confirming such an
inter-change in the model with LM is indeed an artifact of the linearization.

In Figs.~\ref{exp4}(b) we plot the impurity contribution to the
magnetic susceptibility $\chi$.  Dimensionless susceptibility $k T
\chi /(2\mu_B)^2$, where $k$ is the Boltzmann constant and $\mu_B$ the
Bohr magneton, has a peak at intermediate temperatures corresponding
to the local moment regime provided that $g$ is below some critical
value $g_\mathrm{e}$.  For
$g>g_\mathrm{e}$ the local moment regime is absent. By fitting the
Bethe ansatz results for $S=1/2$ Kondo impurity to the numerically
calculated susceptibility, following the procedure described in
Refs.~\onlinecite{desgranges82,rajan82, sacramento89}, we obtain the
estimate of the Kondo temperature $T_K$, shown in Fig.~\ref{exp4}(c).
$T_K$ is increasing rapidly near $g_\mathrm{e}$ as effective
hybridization grows large. For $g>g_\mathrm{e}$ the Kondo temperature is not
defined because there is no local moment in the system.

\begin{figure}
\begin{center}
\includegraphics[width=60mm, keepaspectratio]{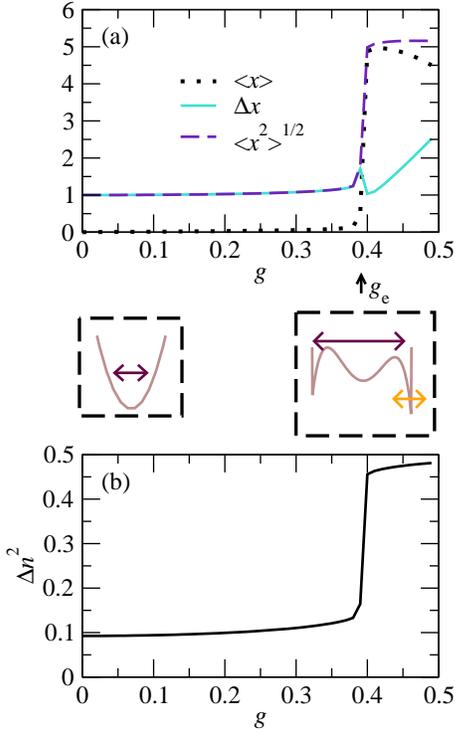}
\end{center}
\caption{\label{exp1} (Color online) (a) Average displacement and its
  fluctuations.  (b) Fluctuations of charge.  Phonon cutoff
  $L=10$, $\zeta=0.1$,  other parameters as in Fig.~\ref{Fig2}. 
}
\end{figure}

In Fig.~\ref{exp1}(a) we plot the averages of displacement and its
fluctuations for $\zeta=0.1$. For $g>g_\mathrm{e}$, the displacement rises
abruptly to the values that are limited only by the phonon cutoff; for
$\zeta=0$ (not shown) the same applies to the fluctuations of
displacement (displacement itself is zero). The abrupt increase is due to the
increased hybridization, which is exponential in displacement. The
gain in the kinetic energy cannot be compensated by the cost in oscillator
potential, which is only quadratic in the displacement operators. In
this regime, due to the exponentially increased hybridization, the
fluctuations of charge, shown in Fig.~\ref{exp1}(b), are near-maximal.
The molecule resides in the effective
potential presented in the right pictogram, which has the same form as
the function plotted in Fig.~\ref{Fig_func}(b) (dashed), and is
attracted to the right electrode. 

\begin{figure}
\begin{center}
\includegraphics[width=60mm, keepaspectratio]{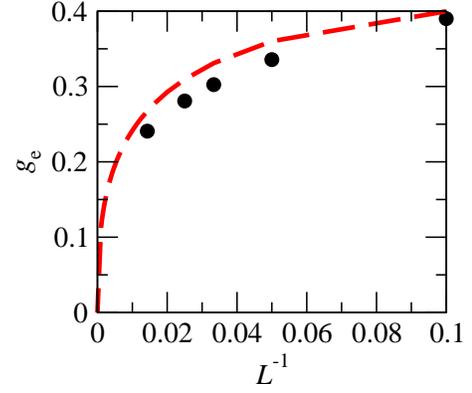}
\end{center}
\caption{\label{exp2} (Color online)
  Values $g_\mathrm{e}$ where the breakdown of the exponential model
  occurs  for phonon 
  cutoffs $L=10, 20, 30, 40, 70$ (circles). Other parameters are as in
  Fig.~\ref{Fig2}.  Semi-classical estimate of $g_\mathrm{e}$ 
  (dashed). 
}
\end{figure}

The value $g_\mathrm{e}$ can be estimated by first recognizing that
the maximum $x$ is bounded by the phonon cutoff $L$:  $x \lesssim 
x_\mathrm{max} =2 \sqrt{L}$ and then solving  $\Omega x_\mathrm{max}^2/4 = V
\exp (g x_\mathrm{max})$ for $g$, which gives
\begin{equation}
  g_\mathrm{e}=\frac{1}{2\sqrt{L}} \log\left(\frac{\Omega}{V} L \right).
\end{equation}
The comparison between this estimate and the numerical data is shown
in Fig.~\ref{exp2}. In the limit of large number of allowed phonons
$L\to\infty$ critical value $g_\mathrm{e}\to 0$. An important result is that the
model with exponential modulation of hybridization is not well-defined
with quadratic stabilizing potential only as the results strongly
depend on the cutoff.

\begin{figure}
\begin{center}
\includegraphics[width=60mm, keepaspectratio]{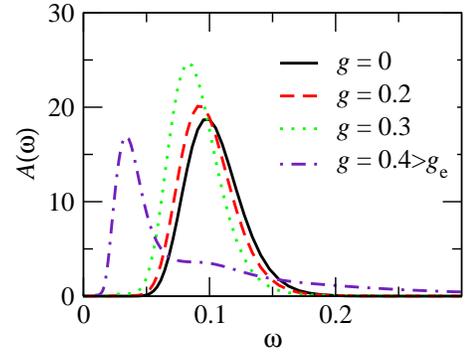}
\end{center}
\caption{\label{exp3} (Color online) Spectral functions exponential
  modulation.  $\zeta=0.1$, $L=10$, other parameters as in
  Fig.~\ref{Fig2}. 
}
\end{figure}

  In Fig.~\ref{exp3} we plot the spectral function $\mathcal{A}(\omega)$. With increasing $g\lesssim
  g_\mathrm{e}$ the bare oscillator peak starts to soften. At
  $g>g_\mathrm{e}$ the molecule is attracted next to the hard wall boundary 
  and remains mainly trapped into one of the wells
  defined by the phonon-cutoff. The high-frequency part of the 
  spectral function in Fig.~\ref{exp3} corresponds to the oscillations within these
  wells. The wells are strongly anharmonic which is reflected
  in the  broad distribution of spectral weight. The low frequency part
  of the spectral weight is due to the oscillations between the wells.

  A natural question which arises at this point is whether it is
  possible to tune the parameters so as to drive EM to the regime
  with developed double well potential, but for $g$ sufficiently lower
  than $g_\mathrm{e}$, so that the wells are not next to the hard wall
  boundary. For the parameter set used in the results shown, for
  example, this is not possible, because
  \begin{equation}
    \label{eq:ineq_last} g_\mathrm{d} >g_\mathrm{e}
  \end{equation}
  for all $L$.

  It can be shown that the inequality Eq.~(\ref{eq:ineq_last}) holds
  in general.  We begin by maximizing $g_\mathrm{e}$ with respect to
  $L$ (treating $L$ as if it was a continuous variable) and obtain
  \begin{equation}
    g'_\mathrm{e} =\max_L g_\mathrm{e}(L)=\frac{\pi^{1/4} \sqrt{\Omega}}{e(D \Gamma)^{1/4}}.
  \end{equation}
  The ratio between $g_\mathrm{d}$ defined in Eq.~(\ref{eq:semiclass})
  and $g'_\mathrm{e}$ can now be evaluated. We obtain that
  $g_\mathrm{d}/g'_\mathrm{e} \geq (e^{7} \pi)^{1/4}/4 \sim 1.92$. The
  double well potential thus develops also in EM but the wells are
  near the boundary defined by the phonon cutoff.

\subsection{Regularized exponential model}
\begin{figure}
\begin{center}
\includegraphics[width=55mm, keepaspectratio]{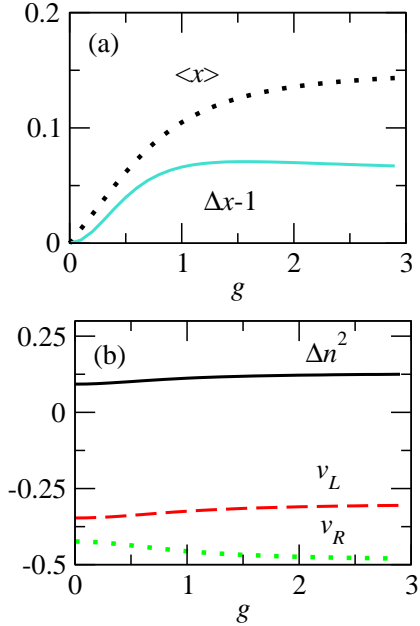}
\end{center}
\caption{\label{th1} (Color online) Results for a model with overlap
  integrals modulated as $V_{L,R} (x) \sim \exp (\mp gx) / \cosh(g
  x)$, Eq.~(\ref{vtanh}), with $\zeta=0.1$ and other parameters as in
  Fig.~\ref{Fig2}. (a) Displacement and its fluctuations.  (b)
  Expectation values of hoppings and fluctuations of charge. }
\end{figure}

By using EM some of the nonphysical results found in the model with LM
are eliminated but others, such as the dependence on a cutoff
parameter $L$, are introduced. Another possibility is to regularize
EM,
\begin{equation} V_{L,R}(x)=V \left[ \exp(\mp gx)/\cosh(gx) \mp \zeta\right], \label{vtanh}
\end{equation}
or in the symmetrized basis \eb V_e=\sqrt{2} V,\: \: V_o =\sqrt{2} V
\left[\tanh(gx)+\zeta\right].\ee The inequality Eq.~(\ref{eq:ineq}) is
satisfied and the normalization with the cosh function ensures the
model behaves well for large $x$.

For this model we evaluated the matrix elements of the Hamiltonian
$\langle m |H| n \rangle$ ($|m\rangle,|n\rangle$ correspond to states
with $m,n$ excited phonons) in the real space by reintroducing the
Hermite polynomials, which is numerically more stable than the
expansion of $\tanh(gx)$ in the power series in $x$. This procedure
can be used for any form of modulation. For EM the procedure can be
simplified because it is possible to evaluate $\langle m | \exp(gx) |n
\rangle$ analytically via the Baker-Hausdorff equality.

\begin{figure}
\begin{center}
\includegraphics[width=55mm, keepaspectratio]{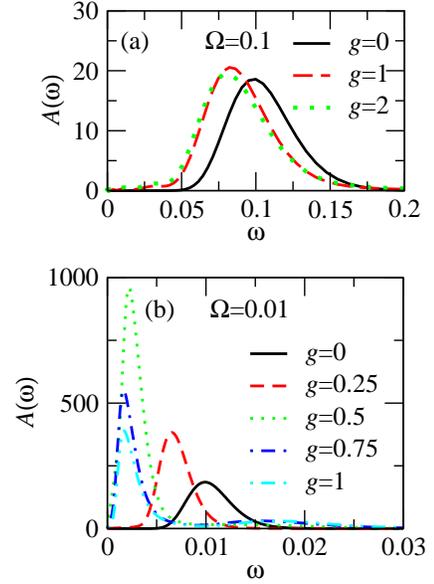}
\end{center}
\caption{\label{th2} (Color online) (a) Displacement spectral
  functions for standard parameters as in Fig.~\ref{th1}. (b) Spectral
  functions for the same set of parameters, except for softer spring
  constant $\Omega=0.01$ and $\zeta=0.01$. }
\end{figure}

We first present results for parameters kept as in LM,
Fig.~\ref{Fig2}, and with asymmetry parameter $\zeta=0.1$.  Due to
weaker dependence of the overlap integrals on displacement the effects
of the electron-phonon coupling in this model are less pronounced. The
displacement and its fluctuations, which we plot in Fig.~\ref{th1}(a)
are small.  This is accompanied by only a minor softening of the
phonon mode, as shown in Fig.~\ref{th2}(a). The fluctuations of the charge
and the expectation values of hopping, plotted in Fig.~\ref{th1}(b)
are likewise only minorly distorted from the $g=0$ case. Note that we
used $\zeta=0.1$ here, therefore the hoppings toward left and right
electrodes differ considerably already for the $g=0$ case, a feature
which is only slightly (compared to LM and EM) amplified by the
dependence of the overlap integrals on $x$ for $g>0$. The reason for
this moderate dependence of quantities on $g$ is the fact that the
largest energy the system can gain by increasing the displacement is
of the order of $V\propto \sqrt{\Gamma}$, which is for these parameters
comparable to the elastic energy $\Omega x^2$ already for the
displacements of order 1.

For smaller $\Omega=0.01$, {\it i.e.}, a softer 'spring', the effect
of the electron-phonon coupling is larger and the soft mode is clearly
developed, Fig.~\ref{th2}(b).  Simultaneously other quantities
($\langle x \rangle$, $\Delta x$, etc.) also exhibit more pronounced
behavior, similar to LM and EM cases (not shown here).

The regularized form, Eq.~\ref{vtanh}, might describe well the
modulation of overlap integrals in realistic case and we believe that
the softening of the phonon mode and related tendency toward broken
symmetry configuration is the universal consequence of
displacement-modulated hybridization.

\section{Conclusion} 

In summary, we analyzed the influence of the electron-phonon coupling
in molecular bridges consisting of a molecule oscillating between two
electrodes.  The overlaps between the molecular orbital and the
orbitals in the electrodes are determined by the position of the
molecule $x$. To model this situation we used several types of the
dependence of the overlap integrals on $x$.

We find that the inversion symmetric model with linear modulation has
a 2CK critical point at some critical electron-phonon coupling
$g=g_\mathrm{c}$ where both channels participate equally to the
screening of the spin. The occurrence of this critical point is
suppressed if a finite difference $\zeta$ between the coupling to left
and right electrodes is introduced. In such broken symmetry system the
sharp transition between the two Fermi liquid states via a non Fermi
liquid state is replaced by a continuous rotation of the screening
state in the channel space from an almost symmetric to an almost
antisymmetric linear combination of the left and right states. This
continuous rotation is accompanied by a dip in conductance near the
point where one of the electrodes is decoupled.

Additionally we found that the electron-phonon coupling modifies the
shape of the effective potential affecting the static and dynamic
properties of the oscillator. At moderate $g$ the potential is
softened and the frequency of oscillations decreases. At large $g$ the
potential develops side-wells and the phonon propagator consists of a
part corresponding to high frequency oscillations within the well and
another part corresponding to slow tunneling between the degenerate
wells.

For finite $\zeta$ the degeneracy between the wells is broken by an
amount $\propto \zeta$ and when the softened frequency $\omega_0$
drops below this value, the tunneling to the higher well is
suppressed.  In this regime, the average displacement starts to
increase significantly and its fluctuations decrease. Hence, only a
minor ($\sim \omega_0$) breaking of inversion symmetry can result in a
significantly asymmetric configuration, which could also account for
the asymmetric configurations typically observed in some experiments.

We consider also overlap integrals exponentially modulated by the
displacement, which is closer to reality in the respect that the
coupling to the even channel is invariably stronger.  However, due to
the exponential increase in hybridization energy gain, this model is
not stable against large distortions. The value of $g$ at which this
instability starts is set by the phonon cutoff and vanishes in the
limit of large cutoff.  We analyzed the model at finite cutoff and
found that the 2CK fixed point is absent, but the other behavior of
the LM is qualitatively recovered. In particular, the softening occurs
with increasing $g$ and for large $g$ the molecule is attracted to one
of the electrodes and is localized next to the boundary given by the
phonon cutoff. In this regime, the Kondo temperature is significantly
increased but the conductance is suppressed due to the small overlap
with one of the electrodes.

The main common finding is thus the softening of the phonon mode,
emerging for linearized and exponential modulations, as well as for
the case of a more realistic regularized modulation.  If the inversion
symmetry is broken, the instability due to the formation of a double
well effective potential will manifest as the attraction of the
molecule into one of the wells and simultaneous suppression of the
conductance. The instability inspired also a very recent work, in
which the break-junctions are studied as an example of a two-level
system\cite{lucignano08}.

Finally, let us comment on the relevance of the presented work outside
the scope of the experiments with molecular conductors. Within the
dynamical mean field theory (DMFT) \cite{georges96} the bulk correlated
electron systems are solved by mapping onto impurity
problems. Likewise, bulk systems with electron-phonon coupling are
mapped onto impurity (or impurity-cluster) problems with coupling to
phonons. In this regard the knowledge of the behavior of the impurity
problems is a convenient guide in the interpretation of the DMFT
results.  The results obtained in this work for the linearized model
correspond to the general two-band case, where the coupling to one of
the bands is phonon-assisted. The large $g$ regime which we dismissed
as unphysical could prove relevant in this context.

Our results could also be applied to the studies of
nanoelectromechanical systems \cite{gorelik98,craighead00,
  cleland02,flowers-jacobs07,doiron08} (NEMS).  In NEMS the tunneling
to electrodes is modulated by the displacement of the cantilever in a
similar fashion as analyzed in this work. Once the dimensions of these
devices are reduced to such an extent that the frequencies of the
oscillations will become comparable to other scales, such as the bias
at which the devices are operated, softening of the vibrational mode
and susceptibility towards large displacements could be observed. For
instance, we predict that the frequency of the oscillations will
decrease if the tunneling rate from the electrodes to the
perpendicularly situated cantilever immersed between them is
increased.

\begin{acknowledgements}  We 
acknowledge  discussions with  T. Rejec and his contributions in the development of the SG code 
as well as discussions with R. \v{Z}itko and the use of his implementation of NRG
(http://nrgljubljana.ijs.si). We thank also P. Prelov\v sek for his inspiring remark.
The work is supported by Slovenian Research  Agency (SRA) under grant Pl-0044.
\end{acknowledgements}

\appendix 
\section{Comparison to Sch\"onhammer-Gunnarsson projection-operator
  method}

\begin{figure}
\begin{center}
\includegraphics[width=60mm, keepaspectratio]{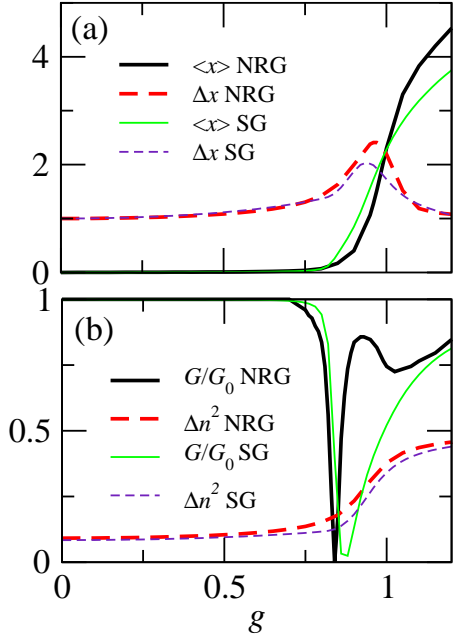}
\end{center}
\caption{\label{Fig_primerjava} (Color online)
Comparison between Sh\"onhammer-Gunnarsson and NRG results. (a)
Displacement and displacement fluctuations. (b) Conductance and charge
fluctuations. $U=0.3$,
$\Gamma=0.02$, $\Omega=0.1$, $\zeta=0.01$. 
}
\end{figure}

In this appendix we compare the results of NRG calculations to the
results obtained by the SG method. Let us summarize first the
results we have obtained using the SG method for model II of
Refs.~\onlinecite{mravlje06, mravlje07}. We have found that for $g$
large enough the variational solution with broken inversion symmetry  is 
lower in energy. This occurs  even for $\zeta=0$ when the inversion
symmetry should persist.  That indicates that for large $g$ 
due to the instability in the system the SG method fails giving
a solution with 'spontaneously' broken symmetry. Such failure is
typical of the mean-field treatment. However, from our previous
analysis\cite{mravlje06} it was not 
clear whether the failure occurs at the point where the 
symmetry of the screening channel is changed  or at the point when the
soft mode is formed (or the two phenomena occur  simultaneously).

In Fig.~\ref{Fig_primerjava} we compare the displacement, its
fluctuations and the fluctuations of charge calculated by SG method to
the NRG results.
They match closely, 
with the exception of  discrepancies in the precise values
of increased displacement fluctuations and range of $g$ where these
occur. Conductance is discussed in the
main text, here we re-plot the curve for completeness.

\begin{figure}
\begin{center}
\includegraphics[width=70mm, keepaspectratio]{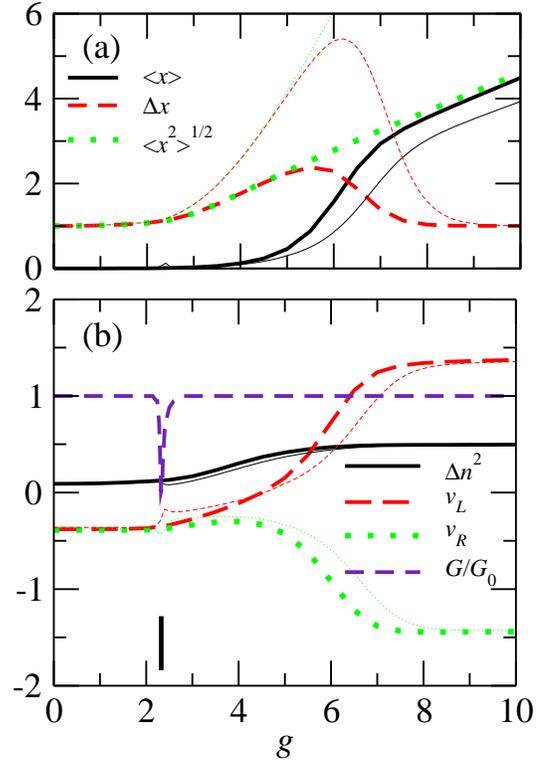}
\end{center}
\caption{\label{omega1} (Color online) $\Omega=1$, other data as in
  Fig.~\ref{Fig_primerjava}. Thick lines: NRG, thin 
  lines: SG; (a) Expectation values of $x$ and its
  fluctuations.  (b) Fluctuations of occupancy and expectation values
  of hoppings to left and right electrodes. The vertical lines denote
  $g_\mathrm{c}$ (thick) and the value of $g$ where the minimum of conductance calculated by SG
 method occurs (dotted). Conductance curves (dashed) obtained by NRG and SG
 coincide.
}
\end{figure}

In Fig.~\ref{omega1}(a) we show the results for $\Omega=1$ where the
discrepancy between NRG results (thick) and SG results (thin) is
larger. In this regime SG method overestimates the value of
displacement and its fluctuations. More interestingly, in
Fig.~\ref{omega1}(b) the jump in expectation values of hopping
operators and a minimum in the conductance are seen in the variational
results near the value $g_\mathrm{c}\sim 2.3$. Here also a small peak
in displacement is seen.

If $\zeta=0$ there is no 'spontaneous' symmetry breaking in SG method
for this $\Omega$. By combining these results we conclude that the
appearance of asymmetric solution in SG will coincide with 
the change in the symmetry of the screening channel from even to odd,
but only when the soft mode is sufficiently developed, otherwise only
a  finite jump in $v_L$, $v_R$ occurs there.

Let us remark here also that in terms of the effective Hamiltonian,
the $g<g_\mathrm{c} (g>g_\mathrm{c})$ regimes 
correspond to the state, where the hopping to left and right electrodes
has equal and opposite phases, respectively. The non-Fermi liquid
$g=g_c$ regime cannot be described in terms of (Fermi-liquid)
effective Hamiltonian. Nevertheless, insisting upon this description,
it can be regarded as a 
combination of states which correspond to two effective Hamiltonians in
which left and right electrodes are, respectively, decoupled.

\section{Schrieffer-Wolff transformation}
To obtain the effective-low energy Hamiltonian $H_\mathrm{eff}$ we first 
divide  the Hamiltonian into two parts\cite{coleman_salerno} 
\begin{equation} H=H_0
+ \lambda H' 
\end{equation} 
where $H_0$, which  for
our example reads
\begin{equation} H_0=\epsilon n +U n_\uparrow n_\downarrow +\Omega
  a^\dagger a + H_L +H_R,
\end{equation}   
is diagonal in the low ($n=1$, no excited phonons) and high ($n=2,0$,
with excited phonons) energy subspaces
(for $\epsilon \sim -U/2$, $U$ and $\Omega$ large).
The hybridization part

\eb H'= V \widehat{v}_{e} + V gx \widehat{v}_{o} \ee 
provides the mixing between the low and high energy
subspaces and $\lambda$ serves as an expansion parameter to be
 set to $1$ at the end of the derivation.  

 Then a canonical transformation generated by some unitary operator $e^{\cal S}$ is
performed to obtain the block-diagonal 
\begin{equation}
\label{eqn:heff}
 \tilde{H}=e^{{\cal S}} H e^{-{\cal S}}  =\begin{pmatrix} H_L & 0 \\ 0 & H_H 
 \end{pmatrix}.
\end{equation}
By expanding Eq.~(\ref{eqn:heff})   
in terms of nested commutators $\tilde{H}= H + [{\cal S},H] + [{\cal S},[{\cal S},H]]/2
+...$ and the generator ${\cal S}$ as a power-series in
$\lambda$, ${\cal S}={\cal S}_1 \lambda +{\cal S}_2 \lambda^2+...$, to second order in
$\lambda$ the following should hold:

\begin{align}
\label{eqn:heff2}
 \tilde{H}= H_0&+ \lambda (H'+ [{\cal S}_1,H_0])  + \\ &+ \lambda^2
 ([{\cal S}_1,H']+\frac{1}{2}[{\cal S}_1,[{\cal S}_1,H_0]]+[{\cal S}_2,H_0]).\nonumber 
\end{align}
Relation Eq.~(\ref{eqn:heff2}) is satisfied to lowest order in $\lambda$ 
by demanding 
\begin{equation} H'+[{\cal S }_1,H_0] = 0.
\label{eqn:lowestorder}
\end{equation}
Since ${\cal S}_1$ can be chosen completely block-off-diagonal the commutator
$[{\cal S}_1,H']$ is block-diagonal, therefore Eq.~(\ref{eqn:heff}) is satisfied
also to second order in $\lambda$ by 
taking ${\cal S}_2=0$. Looking now at the matrix element of
Eq.~(\ref{eqn:lowestorder}) in the basis of
eigenstates of $H_0$ (we will use Latin indices for states in
low-energy subspace,
and Greek indices for states in high-energy subspace), we obtain
${\cal S}_{\alpha b} = H'_{\alpha b} /(E_{\alpha} -E_b) $. 

Finally, the high energy part is neglected by projecting to the low energy subspace 
\begin{equation}
H_{\mathrm{eff}} =P_{\mathrm{low}} H_0 P_{\mathrm{low}} + H^{\mathrm{I}}_\mathrm{eff}
\end{equation} 
with the projector
\begin{equation} 
P_{\mathrm{low}} = |0\rangle \langle 0 \rangle
\left[ n_\uparrow (1-n_\downarrow) + n_\downarrow (1-n_\uparrow)\right]
\end{equation} 
where $|m\rangle$ denotes normalized phonon state with $m$ excited
phonons and the effective interaction $H^\mathrm{I}_\mathrm{eff}$ is
due to the virtual transitions to the high energy states. Its matrix
elements read
\begin{equation} H_{\mathrm{eff}; ab}^{\mathrm{I}} = \frac{1}{2} \sum_{\gamma \in
  \mathrm{high}}  \left (\frac
{H'^\dagger_{a\gamma} H'_{\gamma b}} {E_a -E_\gamma} + \frac
{H'^\dagger_{a\gamma} H'_{\gamma b}} {E_b -E_\gamma} \right).  
\end{equation}

In this case the Hamiltonian can be up to a constant term recast by
using the spin operators to the  form of the 2CK model
\eb H_{\mathrm{eff}}^{\mathrm{I}}=H_\mathrm{2CK}=J_e
\mathbf{S} \cdot \mathbf{s}_e + J_o \mathbf{S} \cdot \mathbf{s}_o, \ee
where the $s_\alpha$ for $\alpha=e,o$ denote the spin densities which
read 
\begin{equation}
\mathbf{S} = \frac{1}{2}\sum_{ss'}d^\dagger_s \boldsymbol{\sigma}_{ss'} d_{s'}
\end{equation}
for $d$-orbital and likewise for orbitals $\alpha$. Here the
components of $\boldsymbol{\sigma}$ are the Pauli matrices. 
The coupling constant to the even channel is  
\begin{equation}  J_e = 2 V^2 \left(
    \frac{1}{-{\epsilon}}  +
  \frac{1}{{\epsilon} +U}\right).\end{equation}
For the odd channel we get \begin{equation} 
J_o=2 V^2 g^2 \left( \frac{1} {-\epsilon + \Omega} +
  \frac{1}{\epsilon+ U + \Omega}\right) 
, \end{equation}
where the $\Omega$ in the denominators occur since high-energy states
involve one excited phonon. 

For the exponential modulation the hybridization parts read 
\begin{align} H'_e=&V \cosh(gx) v_e, \\
  \nonumber H'_o=& V\sinh(gx) v_o.
\end{align}

The low-energy 
Hamiltonian is again that of the 2CK model and the coupling constants
read 

\eb J_e=2 V_s^2 \sum_{m=0}^\infty \left( \frac{\delta_m^e}{-\epsilon +m \Omega}
+ \frac{\delta_m^e}{\epsilon + U + m \Omega} \right), \ee

\eb J_o=2 V_s^2 \sum_{m=0}^\infty \left( \frac{\delta_m^o}{-\epsilon +m \Omega}
+ \frac{\delta_m^o}{\epsilon + U + m \Omega} \right), \ee
where $\delta_m^{e,o}=|\langle 0 |(e^{gx}\pm e^{-gx})/2 |m\rangle|^2= \exp(g^2)
g^{2m}/m!$ for $m$ even (odd) and zero otherwise, respectively.

\section{Conductance}

The linear response conductance for a general interacting system is
given by the Kubo formula\cite{izumida97}
\begin{equation}
\label{eq:kubo} G= \lim_{\omega\to 0} \frac{e^2}{\omega} \mathrm{Im} C_{II}^R(\omega),
\end{equation}
where 
\begin{equation} C_{II}^R(\omega)=-i \lim_{\eta \to 0}\int_0^\infty [I(t),I(0)] e^{i
    (\omega + i \eta) t} dt  
\end{equation} is the retarded current-current correlation
function and the current operators are defined by the time derivative of the
electrons in the electrodes
\begin{equation} I=\frac{\dot{N}_R-\dot{N}_L}{2}. 
\end{equation}
For Fermi liquid systems at $T=0$ the current-current correlation
function can be expressed in terms of the Green's function
$G_{nn'}(\omega)$ involving sites $n,n'$ in the electrodes. The conductance
is then given by the Landauer formula (Fisher-Lee relation\cite{fisher81})
\begin{equation} \label{eq:land} G=G_0 |t(0)|^2,
\end{equation} 
where $G_0=2e^2/h$ is the quantum of conductance and the transmission
amplitude  $t(\omega)$ can be written as\cite{oguri97,oguri01} 
\begin{equation} 
t(\omega)=\frac{1}{-i \pi \rho(\omega)} e^{-ik (n-n')} G_{n'n}
(\omega+i\eta).
\end{equation}

Alternatively, the transmission amplitude can be expressed also in
terms of the scattering matrix $S_{\alpha
  \alpha'}$ which relates
the amplitudes of the outgoing wave in electrode $\alpha$ to the amplitude of the incoming
wave in the electrode $\alpha'$
\begin{equation}
 S=\begin{pmatrix}
r_L & t_R \\
t_L & r_R
\end{pmatrix},
\end{equation}
where $r_\alpha$ ($t_\alpha$) are the respective reflection and transmission
coefficients. The scattering matrix can be rotated in the L-R space to
the basis of channels (linear combinations 
of left and right states) where it is diagonal\cite{pustilnik01}
\begin{equation} 
USU^{-1}=\begin{pmatrix} e^{i 2\delta_a} & 0 \\ 0 & e^{i 2\delta_b} \end{pmatrix}, 
\end{equation}
where $U=\exp(i \theta \tau_y )\exp(i \alpha \tau_z)$ and $\tau_i$ are
the Pauli matrices. According to the Landauer formula Eq.~(\ref{eq:land})
the zero-temperature conductance is determined by the transmission
probability $t(0)=S_{RL}=t_L$, hence we obtain 
\begin{equation}
\label{eq:cond_phase_shift}
G/G_0 = \sin^2 (\delta_a -\delta_b) \sin^2 (2 \theta).
\end{equation}
The angle $\theta$ determines the maximal value of conductance. In the 
inversion symmetric case $\theta=\pi/4$, the phase-shifts
$\delta_\alpha$ which occur in the even and odd channels can then be
extracted from the NRG finite-size spectra\cite{oliveira83}.

When the system is not inversion symmetric it is not possible to
extract $\theta$ since the channel indices are not tracked.  To
evaluate the conductance in this case we 
resort to equation Eq.~(\ref{eq:kubo}). The
current operator  $I=(\dot{N}_L -\dot{N}_R)/2$ is obtained by commuting
$N_{\alpha}=\sum_{k \sigma} n_{k\alpha \sigma}$ with the Hamiltonian
for $\alpha=L,R$ and reads 
\begin{equation}
\dot{N}_{L(R)}=-\frac{i}{\hbar} V_{L(R)} (x) \sum_\sigma (c_{1L(R) \sigma}^\dagger d - h.c.), 
\end{equation}
where $c_{1L(R) \sigma}^\dagger$ creates an electron in the orbital next
to impurity in the left(right) electrode, respectively.

Alternatively, the conductance can be calculated also from the
dependence of the ground state energy on auxiliary magnetic
flux\cite{meden03,molina03,rejec03a,rejec03b} which is obtained by
embedding the interacting system into 
auxiliary noninteracting ring and threading the ring with magnetic
flux $\Phi$. Then the conductance can be expressed
as\cite{rejec03a,rejec03b}  
\begin{equation}
  G/G_0 = \sin^2\left[\frac{\pi}{2} \frac {E(\pi)-E(0)}{\Delta} \right],
\end{equation}
where $E(\Phi)$ is the ground state energy of the auxiliary ring with
embedded interacting system and
$\Delta = 1/\left[\rho(\epsilon_F) N\right]$ the level spacing in the auxiliary
ring with the density of states at the Fermi energy $\rho(\epsilon_F)$
which consists of $N$ sites.  This approach  is useful especially in
connection to variational approaches [such as SG, although in SG we
can extract the conductance also from Eq.~(\ref{eq:land}) directly and
obtain same results], where the ground state energy is the most
reliable quantity. 

\bibliography{oscil}

\end{document}